\documentclass{aastex63}
\usepackage{amsmath,amssymb,multirow,bm}
\usepackage{graphicx}

\renewcommand{\tabcolsep}{1mm}

\newcommand{\be}{\begin{equation}}
\newcommand{\ee}{\end{equation}}
\newcommand{\ba}{\begin{eqnarray}}
\newcommand{\ea}{\end{eqnarray}}
\usepackage{graphicx}
\preprint{}

\begin{document}


\title{Neutron tunneling: A new mechanism to power explosive phenomena in neutron stars, magnetars, and neutron star mergers}

\correspondingauthor{}
\email{carlos.bertulani@tamuc.edu, ronaldo.lobato@tamuc.edu}

\author[0000-0002-4065-6237]{Carlos A. Bertulani}
\affiliation{Department of Physics and Astronomy, Texas A\&M University-Commerce, Commerce, TX 75429, USA}

\author[0000-0001-5755-5363]{Ronaldo V. Lobato}
\affiliation{Department of Physics and Astronomy, Texas A\&M University-Commerce, Commerce, TX 75429, USA}

\date{\today}

\begin{abstract}
Neutron tunneling between neutron-rich nuclei in inhomogeneous dense matter encountered in neutron
star crusts can release enormous energy on a short-timescale to power explosive phenomena in neutron
stars. In this work we clarify aspects of this process that can occur in the outer regions of
neutron stars when oscillations or cataclysmic events increase the ambient density. We use a
time-dependent Hartree-Fock-Bogoliubov formalism to determine the rate of neutron diffusion and find
that large amounts of energy can be released rapidly. The role of nuclear binding, the two-body
interaction and pairing, on the neutron diffusion times is investigated.  We consider a
one-dimensional quantum diffusion model and extend our analysis to study the impact of diffusion in
three-dimensions. We find that these novel neutron transfer reactions can generate energy at the
amount of $\simeq 10^{40}-10^{44}$ ergs under suitable conditions and assumptions.
\end{abstract}


\section{Introduction}\label{sec:intro}
The physics of neutron stars and the role played by the details of strongly
  interacting many-body systems has been a major area of research in astrophysics, based on a
  limited number of astronomical observations that motivated the development of numerous
  theoretical models (for reviews, see, e.g.,
  \citep{BaymPethick1979,Chamel2008,annurev.nucl.50.1.481,Baym_2018,PMID:15105490}).  In particular,
  the physics of neutron stars crusts has attracted the interest of an increasing number of nuclear
  theorists because of, among other phenomena, the prediction of complex structures arising from the
  interactions between nucleons and electrons (see, e.g.,
  \citep{blaschke2008physics,Chamel2008,bertulani2012neutron}). Isospin imbalance occurs in the
  crust of neutron stars where dense neutron regions coexist with neutron poorer regions. However,
  unless disrupting phenomena take place, the crust is rather energetically balanced and little or
  no isospin transfer is expected. If isospin together with energy imbalance is established, the relaxation times leading
  to equilibrium are very short due to beta-decay processes, but neutron tunneling between nuclei
  has also been suggested as a probable reason for fast equilibration in the environment of white
  dwarfs \citep{Saakyan1972}. Similar considerations have been made for the neutron tunneling in the
  crust of neutron stars. In the inner crust, above the neutron drip density of $\sim 4\times
  10^{11}$~{\rm g/cm$^3$} a neutron gas of unbound neutrons exists with all bound states occupied
  thus leading to isospin equilibration by diffusion of unbound neutrons (see, e.g.,
  \citep{Bisnovatyi-Kogan:1979}). In the outer crust, neutron transfer between accreted nuclei has
  been studied with the prediction that they modify the cooling rates in transiently accreting
  neutron stars \citep{Chugunov2018,Chugunov2019}.

A fracture of the crust in a highly magnetized neutron star can reshuffle the magnetic field of the
star. A sudden reorganization of the magnetic field may also be the cause for a crack in the
surface. In either situation, a quick release of stored energy can occur via powerful bursts that
can vibrate the crust, crack it into pieces that move away from each other, a motion that might be
imprinted on gamma-ray bursts and on X-ray signals
\citep{Pacini1974,1989ApJ...343..839B,Huppenkothen_2014}. The breaking strain of the crust by tidal
forces or by resonant elastic modes has also been proposed to generate precursor flares prior to
short gamma-ray bursts due to phase transitions of the lattice, as shown in some theoretical models
\citep{Troja_2010,PhysRevLett.110.011101,PhysRevLett.108.011102}. In
Ref. \citep{PhysRevLett.112.112504}, it was shown that even slight modifications of local neutron
densities in the crust above the neutron drip density can give rise to an attractive interaction
between the nuclei via interstitial neutrons in a lattice formed by nuclei and electrons. This
mechanism likely leads to {agglutination} of nuclei in a form similar to the formation of inhomogeneous
regions in metallic alloys, also known as spinodal decomposition
\citep{PhysRevLett.112.112504}. Huge energy releases, of the order of $\gtrsim 10^{40}$ ergs are
expected to be generated in such cataclysmic scenarios perhaps being responsible for phenomena such
as burst/flaring in soft gamma-rays repeaters (SGRs) and in anomalous X-ray pulsars (AXPs)
(SGRs/AXPs are commonly called magnetars), as well
as fast radio bursts (FRBs).

The outer crust of a neutron star is composed of a lattice of nuclei in a gas of
  moving electrons. As one enters deeper into the inner crust the nuclei become more neutron rich, up
  to a point where neutrons start dripping out of the nuclei. Even deeper in the crust, nuclear
  clusters with exotic shapes will form, due to a competition between the nuclear and Coulomb
  interactions \citep{PhysRevLett.50.2066}. This neutron-rich system is in energetic
  equilibrium but the rupture of the neutron star crust by tidal forces, resonant elastic modes, or
  magnetic field reshuffling, can fuel the formation of a different kind of inhomogeneous neutron
  distribution (defects in the crystalline structure~\citep{kondratyev/2002}, or impurities, which are represented by nuclei whose value of mass or
      charge ($A_i, Z_i$) differs from the nuclei of the background~\citep{deblasio/1998}) away from energy equilibrium, allowing for the sudden diffusion of
  neutrons by tunneling between the neutron-rich region to the region poorer in neutrons, quickly lowering the energy of the system. A fast homogenization of the neutron density
  ensues with a large release of energy. We exploit if this mechanism could be the responsible for
  bursts in SGRs/AXPs and FRBs. Our study is also important in other astrophysical scenarios, e.g.,
  neutron transfer or diffusion in neutron star mergers can also influence the rate at which a
  locally homogeneous density can be achieved. Deformed neutron-rich lattices and gaps are certainly formed during the
  merging process and/or during a fallback mechanism in a core-collapse supernovae. Supernova
  fallback accretion has been intensively studied as possible site for r-process~\citep{fryer/2006}
  and as source of long-duration gamma-ray bursts in newly formed magnetars~\citep{piro/2011,
    metzger/2018a}. We explore the physics of diffusion by tunneling in inhomogeneous neutron media
  considering the flow of individual neutrons as well as neutron-pairs through the nuclear
  mean-field. For simplicity, we assume charge-neutral systems, i.e., pure neutron matter. Our goal
  is to identify general features and possible scaling laws for the diffusion rates that can be used
  to estimate diffusion rates important for cooling properties of neutron stars, and relaxation
  times.

The neutron-rich impurities considered in this work are not typical neutron-rich nuclei accreted at
the surface of neutron stars, as those considered in Ref. \citep{Chugunov2019}. Within the neutron
star crust, and in particular in the inner crust, the proton fraction is expected to be very small,
with the formation of complex and large neutron-rich structures, e.g., identified in semiclassical
Monte-Carlo simulations \citep{PhysRevC.85.015807}.  A long accepted idea is that large and
strangely-looking neutron-rich nuclei, such as ``pasta nuclei", can be formed within the crust
\citep{PhysRevLett.50.2066}. But it has also been shown that such structures might not exist and
that their formation strongly depends on the symmetry energy part of the equation of state of
nuclear matter \citep{PhysRevC.75.015801}. However, with or without pasta nuclei, there is a
consensus in the literature that very neutron-rich structures are part of the neutron star crust
(see e.g., Figure 3 of Ref. \citep{PhysRevC.85.015807}). Such structures are believed to be immersed
within a low density electron and neutron gas. In energetic equilibrium, the neutrons within
{those regions} are confined due to their increased mutual interactions and do not
diffuse to other {neighboring regions}.
Assuming that a cataclysmic event can deform the lattice which composes the crust,
  disrupting the neutron distribution, which leads to neutron tunneling
  between neutron rich regions and to neutron gaps, we propose that the fast tunneling times of loosely bound neutrons can trigger short gamma/X-ray bursts/flaring activities in magnetars and fast radio burst (FRBs) through the liberation of photons in the crust or in the star magnetic field. Beta decay particles in the strong magnetic field move perpendicular to it in quantized Landau levels and the electron-cyclotron energy will be equal to the electron rest-mass energy. In this scenario these particles would also act as a seed for the high energy electromagnetic radiation. The origin of these electromagnetic activities as well as the sources of the FRBs are unknown. Recent observations have shown a connection between these phenomena~\citep{lin/2020, bochenek/2020a, andersen/2020}, maybe solving the puzzling mechanism of FRBs sources.

To understand the physics of neutron tunneling times and how theoretical perturbative and non-perturbative models can be used to obtain realistic estimates, we study a one-dimensional system where a neutron dense region has at least two similar neighbors. One dimensional models, such as the 1D Ising model, have fundamentally impacted our knowledge of
thermodynamics, critical phenomena, particle physics, conformal quantum field theories, magnetism, and emergence in many-body systems. The existence of two neighbors enhances the equilibration rates, and in three-dimensions this enhancement will increase appreciably. Because of resonant tunneling, neutrons diffuse primarily to states with approximately the same single-particle energies followed by decay to states at lower energies or by other nuclear processes such as beta-decay or gamma emission. Transfer of loosely-bound neutrons and the presence of neighbors leads to neutron diffusion estimates that deviate considerably from the perturbative calculations.  As expected, the neutron-neutron interaction and pairing are important effects not amenable to perturbative treatment.

\section{HFB diffusion model.}\label{sec:hfb}
We consider the dynamics of a one dimensional system of neutrons in a
one-body potential $U(x)$ and a neutron-neutron interaction $v(x, x')$ solving the Time-Dependent Hartree-Fock-Bogoliubov (TDHFB) equations~\citep{ringschuck80}
\begin{subequations}\label{hfbtd}
\begin{eqnarray}
i \hbar \frac{\partial}{\partial t} u_{\alpha}(x, t)=\left\{-\frac{\hbar^{2} \Delta_{x}^{(2)}}{2 m(\delta x)^{2}}+U(x)+\Gamma(x)\right\} u_{\alpha}(x, t)
-\delta x \sum_{x^{\prime}} \Delta\left(x, x^{\prime}\right) v_{\alpha}\left(x^{\prime}, t\right)
\end{eqnarray}
\begin{eqnarray}
i \hbar \frac{\partial}{\partial t} v_{\alpha}(x, t)=-\left\{-\frac{\hbar^{2} \Delta_{x}^{(2)}}{2 m(\delta x)^{2}}+U(x)+\Gamma^{*}(x)\right\} v_{\alpha}(x, t)
-\delta x \sum_{x^{\prime}} \Delta^{*}\left(x, x^{\prime}\right) u_{\alpha}\left(x^{\prime}, t\right),
\end{eqnarray}
\end{subequations}
where $\hbar^2/2m=20.73$ MeV fm$^2$, $|u_\alpha|^2$ ($|v_\alpha|^2$) represents the probability that a pair state $\alpha$ is occupied (unoccupied), $\delta x$ is the size step of a discretized one-dimensional mesh, and $\Delta_{x}^{(2)}$ is the second-order differential operator $\Delta_x^{(2)} \phi(x)=\phi(x+\delta x)-2 \phi(x)+\phi(x-\delta x)$. The other quantities are defined as
\ba
  \Gamma(x) &=& \sum_{x^{\prime}} v\left(x-x^{\prime}\right) \rho\left(x^{\prime}, x^{\prime}\right) \\
  \Delta\left(x, x^{\prime}\right) &=& v\left(x-x^{\prime}\right) \kappa\left(x, x^{\prime}\right)\\
  \rho\left(x, x^{\prime}\right) &=& \sum_{\alpha} v_{\alpha}^{*}(x, t) v_{\alpha}\left(x^{\prime}, t\right)
  \\ \kappa\left(x, x^{\prime}\right) &=& \sum_{\alpha} v_{\alpha}^{*}(x, t)
  u_{\alpha}\left(x^{\prime}, t\right),
  \ea
where $\rho\left(x, x^{\prime}\right)$ is the density matrix, $\kappa\left(x, x^{\prime}\right)$ is the pairing density matrix, $ \Delta\left(x, x^{\prime}\right)$ is the pair correlation matrix, and $\Gamma(x)$ is the interaction density. These time-dependent coupled-equations are solved with a fourth-order classical Runge-Kutta method.

The initial ($t=0$) wavefunction is obtained by diagonalizing the standard Hartree-Fock-Bogoliubov (HFB) equations using an expansion of single-particle states in a harmonic oscillator basis with particle-number conservation enforced with the Lagrange multiplier method \citep{ringschuck80}.  This model yields the initial states $\alpha$ and their energies and occupation numbers for a system of $N$ neutrons. We assume spin symmetry, so that $u_{\alpha\uparrow}=u_{\alpha\downarrow}$ ($v_{\alpha\uparrow}=v_{\alpha\downarrow}$) which reduces the working model space to half the number of states needed.

We consider initially neutron-rich impurities of typical nuclear sizes using a confining potential $U_{t=0}(x) = U(x) +U_\lambda(x)$ with
\be
U(x) = -{U_0\over [1+\exp\{(|x|-d)/a\}]},
\ee
and parameters, $U_0 = 100$ MeV, $d=5$ fm and $a=1$ fm. For the neutron-neutron potential we assume a Gaussian interaction of the form
\be
v(x,x')=v_0 \exp\left(-{|x-x'|^2\over 2\sigma_0^2}\right), \label{ve}
\ee
 with $v_0=-14$ MeV and $\sigma_0=2.5$ fm.  To simulate loosely-bound neutrons and obtain the chemical potential close to the continuum, we add to $U(x)$ at $t=0$ a confining harmonic oscillator potential $U_\lambda (x) = \lambda x^2$ and use $\lambda$ as a parameter to adjust the binding energy of the system.

 For $N=20$, 40, 80 and 160 neutrons, with $\lambda = 2$, $10^{-1}$, $1.15\times 10^{-2}$, and $2 \times 10^{-3}$ MeV/fm$^{-2}$, and the potential parameters for $U(x)$ and $v(x, x')$ as listed above, the solution of the static HFB equations yields valence neutrons bound by $S_n=1.55, $0.96, 0.40, and 0.25 MeV for $N=20$, 40, 80 and 160 neutrons, respectively, where $S_n$ denotes the neutron separation energy. This is displayed in Table \ref{table1} together with other energies in the system. The binding energies per neutron are much larger than for a regular nuclear system, however the physics associated with the diffusion rate of the neutrons can be well understood with this model.

\begin{table}[h]
\begin{tabular}{|c|c|c|c|c|c|c|c|c|c|c|}
\tableline\tableline
N & $\lambda$ [MeV/fm$^{-2}$]&$E_{s.p.}$&$E_{kin}$&$E_{int}$&$E_{pair}$& $E_{total}$ & $S_n$ \\
\tableline
20 &$2$& -1353&767.9&-690.4&-6.72	&-1282&	1.55\\
40 &$0.1$& -1830&1384&-1457&-4.97	&-1908&	0.96	\\
80 &$1.15\times 10^{-2}$& -1812&2222&-2550&-7.27	&-2146&	0.40	\\
160 &$2.0\times 10^{-3}$& -1610&3632&-4466&-12.3	&-2456&	0.25	\\
\tableline\tableline
\end{tabular}
\caption{Single particle energy, $E_{s.p.}$, kinetic energy, $E_{kin}$, interaction energy, $E_{int}$, pairing energy, $E_{pair}$, total energy, $E_{total}$ and neutron separation energy, $S_n$, for a system of $N=20$, 40, 80 and 160 neutrons confined in a potential with parameters described in text. All energies are in units of MeV. \label{table1}}
\end{table}

\section{Time evolution and diffusion rates.}\label{sec:TDHFB}
After the preparation of the initial wavefunction for $N$ neutrons, we solve Eqs. \eqref{hfbtd} switching-off the confining potential $U_\lambda$. The potential $U(x)$ as described above, is replaced by a chain of equally-shaped Woods-Saxon(WS)-type potentials separated by a distance $D$. Namely, at $t>0$ we make the replacement
\be
U(x,d,a) \rightarrow \sum_{n=-M}^M U(x-nD,d,a).
\label{Un}
\ee
The initial wavefunction, as described previously, is located at the center of the potential chain, i.e., for the term of the sum with $n=0$. The N-neutron system is thus allowed to evolve with the neutrons tunneling through the barriers with equal widths of about $D-2(d+a)$. The Eqs. \eqref{hfbtd} are solved within a box of size $L=100$ fm, and absorbing boundary conditions with an imaginary potential, located at the edges of the box, with thickness $d_{im}= 50$ fm and strength $W_{im} = -200$ MeV. In the few cases that the box is too small, it was increased beyond $L=100$ fm to accommodate the sequence of $2M$ WS potentials within the box. The absorbing boundary conditions avoid reflections at the borders of the box, relevant for large time scales. The number $2M$ of potential wells entering Eq. \eqref{Un} depends on the distance $D$ between them. Initially, we use $M=2-10$ for $D$ in the range $D=20-100$.

The N-neutron wavefunction is allowed to evolve and the diffusion from the dense to the uncompressed regions predominantly occurs for the neutrons with the smallest separation energies. Tunneling to neighboring region will be partially Pauli-blocked preventing neutrons to flow \citep{Ogata_2020}. Therefore, valence neutrons will be freer to move between the potential pockets and our model captures the relevant aspects of the diffusion process.

The time evolution of the neutron density enable us to calculate the neutron diffusion speed and the
net diffusion coefficient ${\cal D}$ by using the equation $\partial \left< \rho\right>/\partial t =
{\cal D}\partial^2 \left< \rho \right>/\partial x^2$. The diffusion coefficient for the tunneling of
atoms and molecules propagating in a potential lattice, created by a combination of electric,
magnetic, and laser fields in an atomic trap, has been studied, e.g., in
Ref. \citep{BaileyPRA.85.033627}. The diffusion coefficient can be related to relaxation times and
to thermodynamic properties of the system. Despite being a much simpler case than the one considered
here, subtle effects of quantum interference, resonant tunneling, and energy transfer to intrinsic
motion, render a very complicated calculation of the diffusion coefficient
\citep{BaileyPRA.85.033627}.  In our case, we consider a fermionic system which includes additional
microscopic phenomena such as Pauli-blocking and pairing, increasing the degree of difficulty to
calculate the diffusion, or tunneling, rate of the system. Because of that, we will adopt another
calculation procedure, based on the fact that the tunneling rates are very short and amenable to a
simpler approach. In our 1D model we assess the rate at which the neutrons diffuse from a dense to
an uncompressed region by calculating the neutron tunneling rate $\Lambda_n (t)$ from $ \Lambda_n(t)
= - {dN_{0} / dt}$ where $N_{0}(t)$ is the number of neutrons confined within the initial
{neutron-rich impurity} centered at $x=0$ corresponding to $n=0$ in Eq. \eqref{Un}.

Alternatively, the tunneling rate can be calculated using Gamow's model for nuclear decay \citep{Gamow1928}
\begin{eqnarray}
 \Lambda_G = \nu P \sim {v \over d} \exp \left( -2 \int |\kappa (x)| dx \right)
 \sim {\sqrt{2U_0/ m}\over d} \exp \left( -{2\over \hbar} \int \sqrt{2m[U(x)-E]} dx \right), \label{Gamow}
 \end{eqnarray}
where $\nu \sim v/d$ is the barrier assault frequency and $\hbar \kappa = \sqrt{2m[U(x)-E]}$ is the local momentum for a particle with kinetic energy $E$. The integral is performed between the turning points where $E=U(x)$, where $U(x)$ denotes the two connected Woods-Saxon potentials ($n=1$ in Eq. \ref{Un}).  For the most energetic neutrons, i.e., the ones occupying the last orbital for the cases listed in Table \ref{table1}, we have $v \sim \sqrt{2U_0/ m}$ and the Gamow model yields increasing tunneling rates $\Lambda_G \sim (0.5-0.9) \times 10^{-5}$ c/fm for decreasing separation energies $S_n = (1.55 - 0.25)$ MeV.

\begin{figure}[t]
\begin{center}
\includegraphics[
width=3.2in
]
{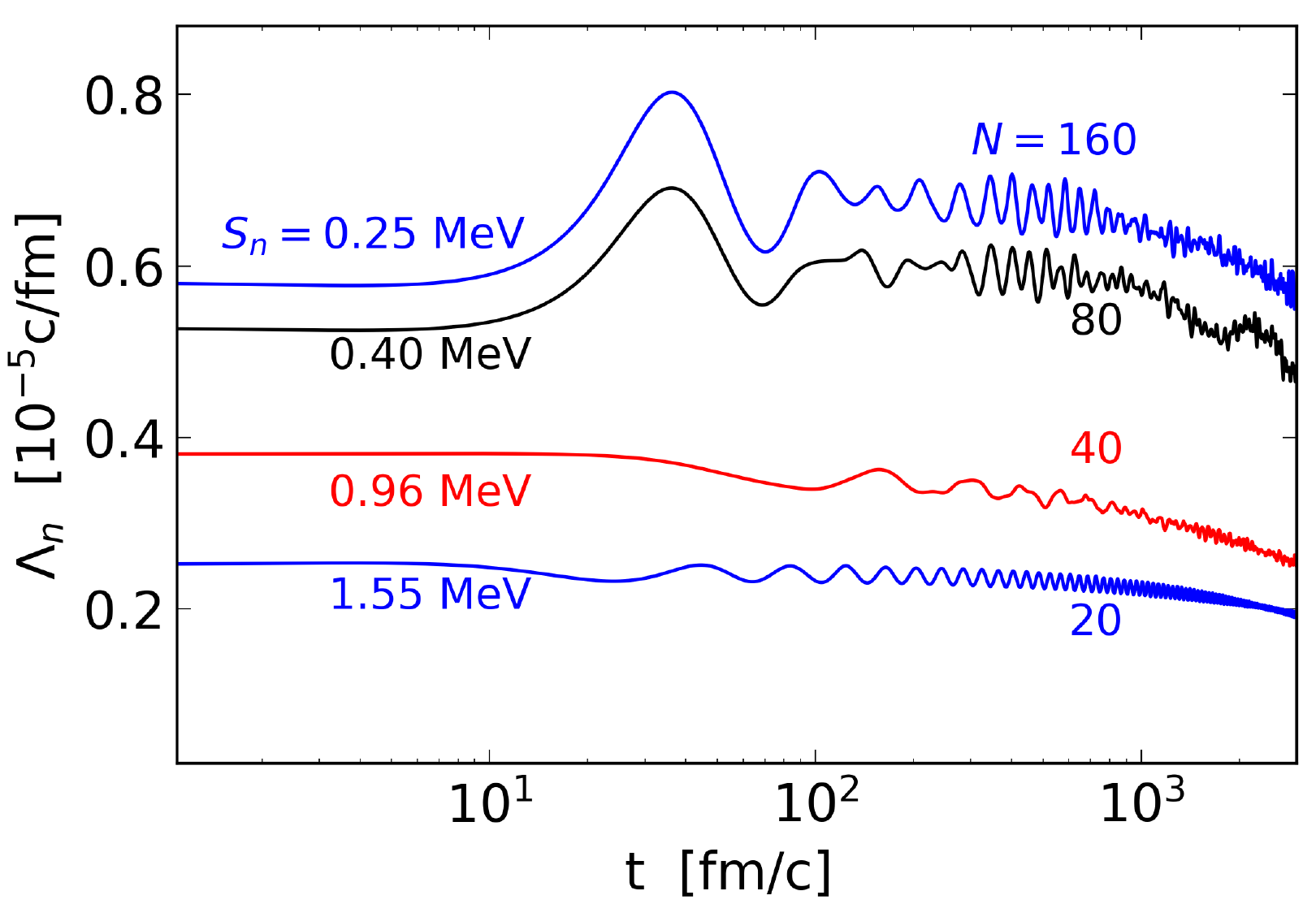}
\caption{Neutron diffusion, or tunneling, rates as a function of time for the neutron-rich impurities separated by 20 fm with neutron numbers listed in Table \ref{table1} and a periodic row of Woods-Saxon potentials as described in the text. }
\label{diffus2}
\end{center}
\end{figure}

As shown in Figure \ref{diffus2}, the neutron tunneling rates $\Lambda_n$ calculated with the dynamical THDFB procedure, and for relatively close region with $D=20$, fm yields values that are not constant in time. Initially, the rates remain approximately constant and are overwhelmingly due to the tunneling of the most energetic neutrons. Oscillations set in at a later stage due to wave mechanical properties such as reflections and interferences. The rates drop at some later time before they flatten out at much larger time scales when less energetic neutrons start participating in the tunneling process. As expected, the neutron tunneling rate increases with decreasing separation energy $S_n$. The rates are smaller by at least an order of magnitude of those predicted by the WKB transmission model.

The smaller transmission rates calculated with the microscopic TDHFB model is partially due to the two-body interaction which makes a large contribution $E_{int}$ to the total nuclear binding, (see table \ref{table1}). An {impurity} is stickier due to the strong neutron-neutron interaction and the ``evaporation" or tunneling to free space regions is suppressed. From table \ref{table1} we see that the contribution of the two-nucleon binding to the total energy reduces from 35 MeV/neutron to 28 MeV/neutron as the neutron number increases for $N=20$ to $N=160$. But the separation energy $S_n$ has a stronger influence on the transfer during the initial stages due to the tunneling of valence neutrons.

\begin{figure}[t]
\begin{center}
\includegraphics[
width=3.2in
]
{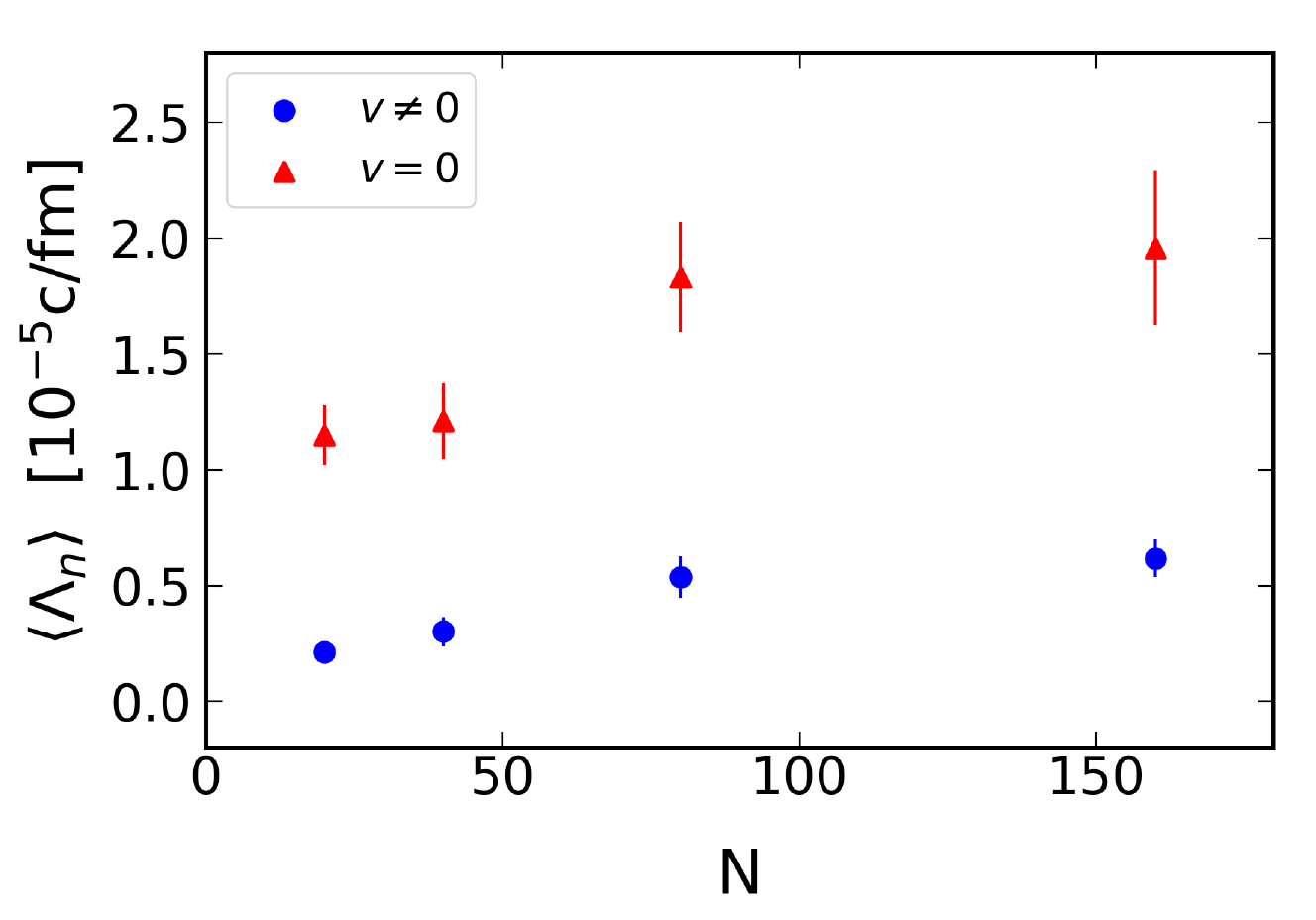}
\caption{Average value of $\left< \Lambda_n \right>$ for $D=20$ fm during the interval $t=0-1000$ fm/c for $N=20$, 40, 80 and 160. The error bars represent the standard deviation of the average values. The diamonds (filled circles) are obtained solving Eqs. \eqref{hfbtd} with the neutron-neutron interaction $v(x,x')=0$ ($v(x,x')\neq 0$). }
\label{diffus3}
\end{center}
\end{figure}

To gain more insight we switch off the two-body interaction $v(x,x')$ at $t>0$. Figure \ref{diffus3} shows the average value of $\left< \Lambda_n \right>$ for $D=20$ fm during the interval $t=0-1000$ fm/c. The error bars represent the standard deviation of the average values. The diamonds (filled circles) are obtained solving Eqs. \eqref{hfbtd} with $v(x,x')=0$ ($v(x,x')\neq 0$). The tunneling rate increases by a factor $\simeq 4-5$ when the neutron-neutron interaction is turned off. This reassures the need of a many-body calculation for the diffusion process when residual nucleon-nucleon interactions are relevant.

Pairing correlations are important in two neutron transfer reactions between nuclei \citep{Bes1969,Broglia1968}. The enhancement of tunneling emerges in transparent analytical models for Cooper pairs and composite particles, as shown in Refs.  \citep{Flambaum2005,BertulaniJPG2007}. We study the impact of pairing switching off the pairing density matrix $\kappa \left(x, x^{\prime}\right)$ in the TDHFB equations \ref{hfbtd}, equivalent to solving the TD-Hartree-Fock equations. The $\Lambda_n$ rates are barely changed, decreasing the tunneling rate by less than 3\% for $N=160$ and less than 5\% for $N=20$. Therefore, there is dominance of single neutron tunneling in the 1D model. There is no direct correlation of this process with neutron transfer in heavy-ion exchange reactions because in the later case there are two time-scales; one for the reaction time, and another for the neutron tunneling dynamics. In our model, pairing is unlikely to modify the total energy by suppressing single-neutron transfer in favor of pair transfer.

We now discuss the tunneling rate dependence on the distance between the neutron-rich impurities and
the neutron gaps. The same physical properties reported above are also observed with increasing separations $D$. The computing time increases considerably because of exponentially smaller tunneling probabilities with increasing separation distances. This feature is displayed in Table \ref{table2} for $N=160$ using the same parameters as in Table \ref{table1} for the initial wavefunction. The diffusion, or tunneling, rates obtained with the TDHFB calculations are about 2-15 times smaller than the predictions based on the Gamow model.

\begin{table}[h]
\begin{tabular}{|c|c|c|c|c|c|c|c|c|c|c|}
\tableline\tableline
D & 20 fm& 35 fm & 50 fm & 100 fm \\
\tableline
$\left< \Lambda_n \right>$ & $5.16\times 10^{-6}$ & $1.13\times 10^{-9}$& $2.24\times 10^{-12}$& $4.77 \times 10^{-24}$\\
$\Lambda_G$ & $ 9.51 \times 10^{-6}$& $1.24\times 10^{-8}$& $1.35\times 10^{-11}$&$2.93 \times 10^{-22}$	\\
\tableline\tableline
\end{tabular}
\caption{Average tunneling rates $\left< \Lambda_n \right>$ (in units of c/fm) with increasing separation distances for $N=160$ using  the same parameters as in Table \ref{table1} for the initial wavefunction. The second row displays the results of the WKB method based on Eq. \eqref{Gamow}. \label{table2}}
\end{table}

The tunneling rates reported here are large compared to typical ones in nuclear reactions because of
the large neutron numbers and small separation energies we have adopted for the valence
neutrons. Because of the Coulomb barriers and symmetry energies in normal nuclei, and for large
separation between them, the tunneling rates are much smaller. On the other hand, as the neutron
number increases in the envelope and crust of neutron stars, the reaction rates are expected to
increase accordingly when inhomogeneous conditions develop. We have also considered neutron-rich
impurities with 500 and 1000 neutrons with widths and depths of the confining potentials $U(x)$
adjusted to accommodate all neutrons within the potential wells while keeping the valence neutrons
at about 1 MeV binding. This time, the distance between the edges of the impurities were kept fixed
at 100 fm. We obtain $\left< \Lambda_n \right> \sim 7.23 \times 10^{-22}$ c/fm and $\left< \Lambda_n
\right> \sim 5.14 \times 10^{-20}$ c/fm for N = 500 and 1000, respectively. This is in agreement
with the increase of the diffusion rate with the neutron number, but it is manifestly stronger for
large {neutron-rich impurities}. It is a probable scenario in a cataclysmic event, i.e., large neutron-rich impurities separated by large distances. If the distances become smaller due to compression waves, the neutron diffusion process can release enormous amounts of energy.

\section{FRBs and magnetars.}\label{sec:FRBs}
Fast radio bursts (FRBs) are a new astrophysical electromagnetic phenomenon discovered in
  recent years. These are radio pulses with unknown origin and the research in this field is fairly nascent. Theories to explain this phenomenon are
diverse \citep{platts/2019}, ranging from highly speculative (as for example, invoking alien
civilizations) to more standard ones, such as merger of compact stars~\citep{totani/2013, liu/2018a} and fracturing crusts~\citep{suvorov/2019}.
The radio pulses are very bright with brief durations, typically in
a range from $\sim30~\mu$s to $\sim20~$ms~\citep{gajjar/2018, michilli/2018, katz/2018} and apparently there is no indication of repetition for the majority of them. Relativistic particle beams with large Lorentz factors $\gamma$ are possibly involved in
the emission process, in a pulsar-like mechanism, where $e^{\pm}$ pairs are created and accelerated
to ultra-relativistic speeds in the polar cap region. Radiation coherence makes $N$ particles radiate with $N^2$ times the
single-particle emission~\citep{cordes/2019}. In a pulsar model the spindown power is responsible for
the electromagnetic radiated power, i.e., the loss of rotational energy of the star provides the
power, therefore an instantaneous emission cannot exceed the spindown power in these radio
emitters. However, this is possible for magnetars (SGRs/AXPs), where the
radiated power is believed to come from the huge magnetic field ($B\sim10^{12}-10^{15}$~G) instead
of rotation, i.e., the decay of the ultra-strong magnetic field generates the
emission. Unpredictable and unknown instabilities in these sources are responsible for bursting/flaring activities in X and gamma-ray spectrum from
few milliseconds to tens of seconds. There are three kinds of bursts: the short ones, with
$\sim10^{39}-10^{41}$~erg/s; the intermediates ones, with $\sim10^{41}-10^{43}$~erg/s; and the giant
flares, which are exceptionally rare events with energies of $\sim10^{44}-10^{47}$~erg/s. According to the McGill~\citep{olausen/2014} online
catalog~\footnote{\url{http://www.physics.mcgill.ca/~pulsar/magnetar/main.html}} only
five of thirty sources are radio-emitters (in a quiescent state), and it seems that the origin of the
radio emission in these sources is different from standard
radio-pulsars~\citep{turolla/2015}.

Recently, it was raised the possibility that FRBs have their origin in
magnetars~\citep{margalit/2018}, and there are some evidences showing that; polarization measurements suggest that FRBs sources are strongly
magnetized, the localization of several FRBs to star-forming regions typical of magnetars~\citep{bochenek/2020}, the soft
gamma-rays repeaters (SGRs) emit giants flares/bursts with volatility, and a more recently
observation~\citep{mereghetti/2020} showed that X-ray bursts from the magnetar SGR 1935+2154 were
also accompanied by a very bright millisecond radio burst.
Several magnetohydrodynamics instabilities can occur in few seconds~\citep{kokkotas/2014}, the lack
of correlation between bursts and waiting times suggests that the trigger mechanism could be
small scale intrinsic (non-global) mechanism and episodic~\citep{suvorov/2019}. \cite{li/2019a} showed that the waiting time ($10^{-2}-10^{-3}$) is of the order of the Alfv\'en crossing time
\begin{equation}
t_{\mathrm{A}} \sim 10^{-3}\left(\frac{\rho}{10^{13}~\mathrm{g}~\mathrm{cm}^{-3}}\right)^{1 / 2}\left(\frac{L}{10^{5}~\mathrm{cm}}\right)\left(\frac{B}{10^{15}~\mathrm{G}}\right)^{-1} \mathrm{s},
\end{equation}
where $\rho$ is the crustal density, $L\sim R_{\odot}-R_c$, with $R_{\odot}$ being the stellar
radius and $R_c$ the crustal radius.

It was suggested that radio emission bursts might originate from the closed field zone within the
near magnetosphere of the magnetar~\citep{wadiasingh/2019}, in a pulsar-like mechanism occurring near the
surface of the star generated by a crust yielding event. Along the same lines, others mechanisms~\citep{beloborodov/2017}
have also considered a crustal/quake event~\citep{petroff/2019}.

Here we propose that the trigger for the
short gamma/X-ray bursts and FRBs could be the diffusion of neutrons in an inhomogeneous density
environment in the crust. The diffusion can occur in a short time in a region at the crust where
neutron-rich nuclei or neutron-rich impurities donate
neutrons to a neighborhood poor in neutron content, generating beta decay or gamma
emission and thus releasing a large amount of energy. It should be stressed, that
  the formation of neutron-rich regions with the described properties is a basic assumption of our model. The electron from the beta decay is relativistic in the regions where the density is
$>10^6~\mathrm{g/cm^3}$ and will act as seeds for the high energy emission and coherent emission which
leads to a brief radio emission. These electrons come out from the crust to the inner magnetosphere and will give rise to a
$e^{\pm}$ cascade,
producing the high-energy radiation via synchrotron or inverse Compton
scattering. These mechanisms will drive Alfv\'en waves, and as shown in Ref. \citep{kumar/2020}, a large-amplitude Alfv\'en wave packet is possible to be
launched by a disturbance in the near surface of the magnetar and part of the wave energy is
converted to coherent radio emission in few tens of the neutron star radii. In our
proposed mechanism, as the magnetic structures evolve in a neutron star, the magnetic field flux tubes passing
from the core to the crust build up a large stress~\citep{ruderman/1998}. This happens for years until the shear strain reaches a critical value $s_{\rm cr}\approx 0.1$ in the lattice~\citep{horowitz/2009}. By exceeding this value, plastic failures are triggered according to molecular dynamics
simulations~\citep{horowitz/2009, chugunov/2010}. These deformations will give rise to the manifestation of a rapidly-acting
hydrodynamics instability~\citep{thompson/1996, rheinhardt/2002}, leading to the emergence of large
amplitudes compressing waves whose part of the energy is promptly dissipated and
  converted into radiation, and part stored in highly compressed waves. The fate of the energy carried by those
  waves are not well known or how they are damped. We propose that this energy is responsible for
  forming inhomogeneous neutron distributions. As the wave
propagates in the crust with a velocity $v\sim 10^{-2}c$~\citep{li/2015}, it interacts with the
solid lattice, donating energy to it, becoming damped in the process, changing the
local density, and forming the neutron-rich impurities which are quickly relaxed by
neutron tunneling, e.g., see the {lower panel} of Fig. 2 of the paper of~\cite{horowitz/2009}, where it is shown in
  red colors the plastic deformations, i.e., where the ion lattice suffers deviations from the ideally
  uniformly sheared {\it bbc} lattice. We assume that those red regions are the place where
 the local changes
in the density occur, leading to the formation of neutron-rich regions, at close distances $L$, together with neighborhoods of poor
neutron content, which can accept neutrons, thus leading to a quick relaxation by neutron tunneling. According to our estimates, the
tunneling process can suddenly release energy and trigger the burst/flaring
in SGRs/AXPs and FRBs. The tunneling process effectively takes no time, and the duration of the burst is related on the time which the wave cross the crust. Changes in the neutron
density could also lead to pressure perturbations $\Delta{P}\sim n_e\Delta{\mu}$, where
$\Delta{\mu}=\mu_e+\mu_p+\mu_n$, i.e., the pressure gradient is related to the chemical
potentials. The pressure gradient leads to perturbations in the magnetic field and
through ambipolar diffusion it heats up the crust~\citep{beloborodov/2016}.

To provide a crude estimate of the total energy released in a burst, $E_{\rm burst}$, we assume that on average the separation energy of the valence neutrons is $\epsilon \sim 1$ MeV and the ensuing tunneling to a neighboring site releases a similar amount of energy.  We assume that
\begin{equation}\label{burst}
E_{\rm burst} = \int  \epsilon N_{\rm imp} \left( 1- e^{-\Lambda_{\rm G}t}\right) dt \sim \epsilon N_{\rm cl}
\Lambda_{\rm G} \Delta t \sim \frac{\epsilon N_{\rm
    cl} 10^2L}{\alpha D} \sqrt{\frac{2U_{0}}{mc^2}}\exp\left(-{2D\sqrt{2mc^2\epsilon} \over \hbar c} \right),
\end{equation}
where $\Lambda_{\rm G}$ is the tunneling rate from Eq. \eqref{Gamow}. The time during which the
tunneling process is effective as the wave sweeps through the crust is given by $\Delta t\approx L/v
= 10^{2}L/c$, with $L\approx 2$ km being the approximate distance that the wave propagates and
dampens by formation of neutron-rich impurities,
and $N_{\rm imp} $ is the number of impurities involved in the process. We further
assume that the impurities
have a dimension $\alpha D$, where $\alpha > 1$ takes into account large {impurity} sizes. Notice that
the separation distance $D$ is now representing the distance between the edges of the neutron-rich
impurities. In Eq. \eqref{burst} we have assumed that $\Lambda_{\rm G} \Delta t \ll 1$. From the numbers
presented in Table 2, this is likely the case for $D \gtrsim 50$ fm. At smaller distances, the
tunneling rates are so large that $\Lambda_{\rm G} \Delta t \gg 1$ and a better estimate is  $E_{\rm
  burst} \sim  \epsilon N_{\rm cl}$. We will use the approximation on the right-hand side of
Eq. \eqref{burst} as a basis for our predictions.

\begin{figure}[!h]
\begin{center}
\includegraphics[
width=3.3in
]
{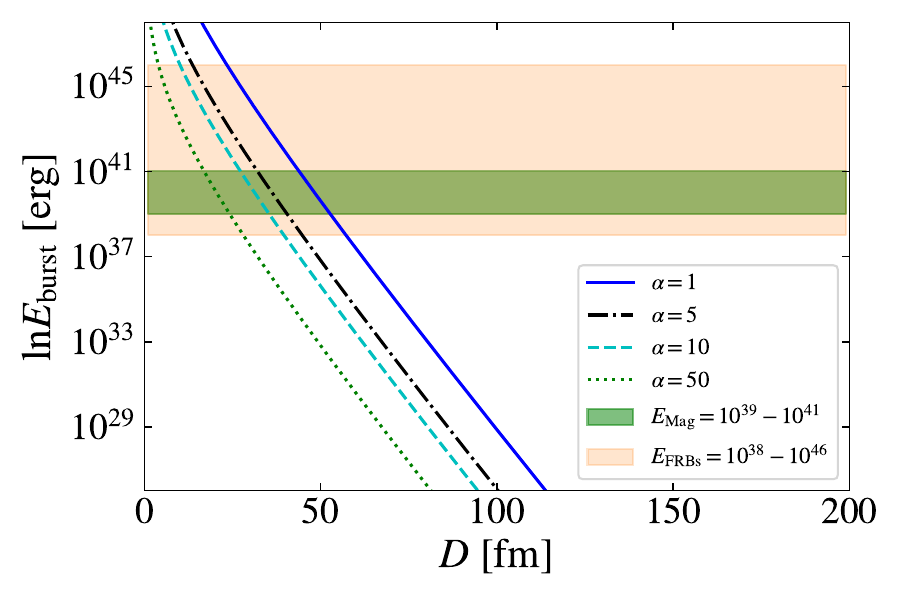}
\caption{Energy burst as function of the separation distance $D$ for $\alpha=1, 5, 10, 50$. The
  green shaded horizontal region represents the observed values of short burst in magnetars, while
  the orange represents the observed values of FRBs.}
\label{energy}
\end{center}
\end{figure}

To determine the number of impurities involved in a rapid emission, we consider
  the polar cap region, which can be estimated through the cylindrical region of radius $R_L$
  encapsulating the closed magnetic field lines. The maximum velocity of the particles within this
  region will be the speed of light, so that $c = \Omega R_L$, with $\Omega$ being the star
  rotational angular frequency. In the particle emission region, the lines are open and the
  boundaries define the polar cap with a radius, in the dipole field, given by $R_{\rm p} = R\sin\theta_{\rm p} = R\sqrt{R/R_L} = R\sqrt{\Omega R/c}$,
  where $R$ is the star radius (see, e.g., Ref.~\citep{ghosh/2007}). Therefore, the polar cap radius will depend on the period $P=2\pi/\Omega$ (for magnetars, $2-12$~s)
and the star radius $R$. Considering a neutron star with $R=11$~km, then $R_{\rm p}\approx 118-48$~m
for $P=2$ and 12 s respectively. The polar cap in a twisted magnetic field configuration is $R_{\rm p} =
R\sin\theta_{\rm p}= R\sqrt{(R\Omega/c)^n/(15+17n)/32}$, where $n$ evolves from $n<1$ to $n=1$. Twisted field lines may enhance the
polar cap radius in magnetars to about 1 or 2 km~\citep{tong/2019}. We can estimate the height under the polar cap, $h_{\rho_0}$, from
the star surface to the layer of density $\rho=\rho_0$, where $\rho_0$ is the nuclear matter
saturation density, yielding $h_{\rho_0}\approx 2$~km. The crust is composed of the outer and inner
crusts: The first one extends from the atmosphere bottom to the layer of density $\rho_{\rm
  ND}=4\times 10^{11}\ \rm{g/cm^3}$ with some hundred meters; the second one is about one km and
the density goes from $\rho_{\rm ND}$ to 0.5$\rho_0$, so one can estimate that the thickness of the
crust is about 1-2~km~\citep{haensel/2007}. The volume of interest for the emission region in magnetars is $V = \pi R_{\rm p}^2h_{\rho_o}$. The number of impurities in this region is
  \begin{equation}
  N_{\rm imp} = {V\over 4/3 \pi (\alpha D)^3}\sim {3R^2 h_{\rho_o} \over 4\alpha^3 D^3}\frac{(R\Omega/c)^n}{(15+17n)/32}.
\end{equation}
We consider three scenarios with $\alpha = 1$, 5, 10 and 50, taking into account
neutron-rich impurities which can also be much larger than the distance between them. In figure~\ref{energy} we show
the energy emitted in a burst as a function of their separation distance $D$, considering a polar cap
radius $R_{\rm p} = 2$ km. The green shaded horizontal region displays the observed values of short bursts in magnetars $\sim 10^{39}-10^{41}~{\rm
  erg/s}$~\citep{turolla/2015} and the shaded orange, the values of FRBs $\sim 10^{38}-10^{46}~{\rm
  erg/s}$~\citep{zhang/2020a}.

Figure \ref{energy} also shows that the bursts can become comparable with the observed values of
short burst in magnetars, and the observed values of FRBs, if the dense neutron sites are within
distances of $D\sim 50$ fm, or below. These are rough estimates based on the WKB approximation with possible corrections by at least an order of magnitude. As we have discussed previously, a microscopic calculation can change these results appreciably if carried out within a proper three-dimensional lattice which offers more tunneling opportunities. One also sees from the figure that smaller impurities favor larger energy yields. This feature is likely to remain as a robust result in more detailed microscopic calculations.

\section{Conclusions}
 We have developed a microscopic TDHFB model to describe neutron diffusion rates due to tunneling
 from neutron-rich impurities to regions voided of neutrons. The model is one dimensional but displays many features of the time dependent behavior of strong interacting particles.

 Our main findings in this study include, but are not limited to: (a) There are marked differences between estimates based on WKB models and the time-dependent microscopic modeling for the detachment and diffusion of neutrons in an inhomogeneous neutron environment. (b) Tunneling is smaller than those obtained with perturbative (i.e., WKB) predictions, but this could change in three-dimensional calculations. (c) The role of pairing is subtle and might strongly depend on the system being studied.

 Time-dependent microscopic calculations show that the subject of density homogenization and isospin diffusion in nuclear reactions and in stellar environments deserves more extensive studies.  Perturbative estimates are likely to yield poor results because of the microscopic properties of strongly interacting systems such as the different contributions of the interactions to the total energy and the related energy rearrangement due to tunneling. Microscopic calculations are rich in physics details and are now becoming feasible for 3D calculations, e.g. Ref.~\citep{StetcuPRL.114.012701} with the shortcoming of costly computation time even with supercomputers. It is worthwhile mentioning that neutron tunneling has been considered in previous works in the context of diffusion of unbound neutrons in the inner crust \citep{Bisnovatyi-Kogan:1979}) and between nuclei in transiently accreting neutron stars \citep{Chugunov2018,Chugunov2019}.

 Non-perturbative calculations of neutron tunneling rates in inhomogeneous neutron distributions may
 play an important role in many astrophysics scenarios including rare events involving neutron
 stars. As proposed in this work, sudden medium modifications caused in cataclysmic environments
 such as supernovae, neutron star mergers, or the creation of energy stored in inhomogeneous regions
 in the crust of neutron stars due to compression waves from a star quake, can all lead to gamma or
 radio bursts when proper conditions for neutron diffusion is attained.  We conclude from our
 analysis, that many-body nuclear physics dictates that such conditions depend on the existence of
 inhomogeneous neutron distributions, i.e., neutron-rich impurities separated by relatively small distances.

\section*{Acknowledgements}
We have benefited form useful discussions with Sanjay Reddy and Takashi Nakatsukasa. This work has been supported in part by the U.S. DOE Grant No. DE-FG02-08ER41533.

\bibliographystyle{aasjournal}

\end{document}